\documentclass{elsart}

\usepackage{natbib}

\usepackage{graphicx}

\usepackage{amssymb}

\usepackage{color}
\newcommand{\Piz}{\ensuremath{\pi^0}}

\newcommand{\Egam}{\ensuremath{E_\gamma}}

\newcommand{\Rpipi}{\ensuremath{\gamma p \rightarrow \Piz{} \Piz{} p}}
\newcommand{\Roper}{P\ensuremath{_{11}}(1440)}

\begin{document}

\begin{frontmatter}


\title{Double $\Piz$ Photoproduction off the Proton at Threshold}
\author[label1]{M.~Kotulla}
\ead{Martin.Kotulla@unibas.ch}
\author[label2]{J.~Ahrens}
\author[label3]{J.R.M.~Annand}
\author[label2]{R.~Beck}
\author[label2]{D.~Hornidge}
\author[label4]{S.~Janssen}
\author[label1]{B.~Krusche}
\author[label3]{J.C.~McGeorge}
\author[label3]{I.J.D.~MacGregor}
\author[label5]{J.G.~Messchendorp}
\author[label4]{V.~Metag}
\author[label4]{R.~Novotny}
\author[label4]{M.~Pfeiffer}
\author[label3]{R.O.~Owens}
\author[label2]{M.~Rost}
\author[label4]{S.~Schadmand}
\author[label3]{D.P.~Watts}

\title{}

\address[label1]{Department of Physics and Astronomy,
      University of Basel, CH-4056 Basel (Switzerland)}
\address[label2]{Institut f\"ur Kernphysik,
  Johannes-Gutenberg-Universit\"at Mainz, D--55099 Mainz, (Germany)}
\address[label3]{Department of Physics and Astronomy, University of Glasgow,
      Glasgow G12 8QQ, (UK)}
\address[label4]{II. Physikalisches Institut, Universit\"at Gie{\ss}en,
      D--35392 Gie{\ss}en, (Germany)}
\address[label5]{KVI, Zernikelaan 25, 9747 AA
      Groningen, (the Netherlands)}

\author{}

\address{}

\begin{abstract}
The reaction $\gamma p \rightarrow \pi^0 \pi^0 p$ has been measured 
using the TAPS BaF$_2$ calorimeter at the tagged photon facility of the
Mainz Microtron accelerator.
Chiral perturbation theory (ChPT)
predicts that close to threshold this channel is
significantly enhanced compared to double pion final states with
charged pions. In contrast to other reaction channels,
the lower order tree terms are strongly
suppressed in 2\Piz{} photoproduction.
The consequence is the dominance of pion loops in the
2\Piz{} channel close to threshold - a result that  
opens new prospects for the test of 
ChPT and in particular its inherent loop terms. 
The present
measurement is the first which is sensitive enough for a conclusive 
comparison with the ChPT calculation and is in agreement with its prediction.
The data also show good agreement with a calculation in the unitary chiral 
approach. 
\end{abstract}

\begin{keyword}

\PACS 11.30Rd \sep 13.60Le \sep 25.20Lj

\end{keyword}

\end{frontmatter}

\section{Introduction}
\label{intro}

In the energy regime where excitations of the nucleon or properties of
the lowest lying mesons are studied, the perturbative ansatz to solve QCD 
fails, because the
strong coupling constant $\alpha_s$ is too large. A different
approach
exploiting the approximate Goldstone boson nature of the pion has been
developed, namely chiral perturbation theory (ChPT) \cite{weinberg:chpt,gasser:chpt}.
This effective field theory
has been extended to the nucleon sector (HBChPT\footnote{ChPT is used in
this paper as a synonym for HBChPT}) 
\cite{jenkins:hbchpt,bernard:hbchpt}. Historically, it was a great success
that ChPT could clarify the
disagreement between the old low-energy theorems (LETs) in describing the
s-wave threshold amplitude $E_{0+}$ in the reaction $\gamma p \rightarrow
\Piz{} p$ \cite{mazzucato:pi0,beck:pi0}.
Additional contributions due to pion loops had to be added to the
old LETs, resulting in a lower value for the 
$E_{0+}$ amplitude and agreement with measured data \cite{bernard:pi0}. 
In general, ChPT is 
in good agreement with experiments describing
$\pi - N$ scattering \cite{fettes:chiral}.

From the study of $\pi \pi$ production processes, complementary
information to the study of the single pion photoproduction 
channels can be gained.
The extention of ChPT to $\pi \pi$ photo- and
electroproduction has led to the finding that the cross section for final
states with two neutral pions is dramatically enhanced due to chiral (pion)
loops \cite{bernard:pipithres} which appear in leading (non vanishing) 
order $q^3$.
This is a counter-intuitive result, since
in the case of single pion production the cross sections for charged pions
are considerably larger than those with neutral pions in the final state.
This situation is not changed when the ChPT calculation is extended 
by evaluating all next-to-leading order
terms up to order $M_\pi^2$ in the threshold amplitude \cite{bernard:2pi0thres}.
Exploring this situation in more detail,
the two pion channel exhibts the following
properties \cite{bernard:2pi0thres}: Born type contributions start at order
$M_\pi$ and are very small. Tree diagrams up to order $q^3$ are
zero due to threshold selection rules or pairwise cancellation. Only at
order $q^4$ do tree terms proportional to the low energy constants $c_i$ give a
moderate contribution. In a microscopic picture these tree terms 
($\varpropto c_i$)
subsume all s-, t- and u-channel resonances in the $\Piz{} p$ scattering
amplitude (e.g. the $\Delta (1232)$). The largest resonance contribution at
order $M_\pi^2$ comes from the \Roper{} resonance via the $N^*N\pi\pi$
s-wave vertex.
Possible double $\Delta$ graphs as
well as loop diagrams containing a photon coupling to a 
$K^+ - \Sigma / \Lambda$ pair were estimated and found to be negligible.
Fourth order loop diagrams ($q^4$) provide only a moderate contribution. 
All the coefficients of the resulting threshold amplitude were taken from
the literature, $\pi - N$ scattering and, in case of the s-wave \Roper{} to
$\pi\pi$ coupling, from an analysis of the 
reaction $\pi N \rightarrow \pi \pi N$ \cite{bernard:roper}. 
Adding all contributions together, the astonishing result is that the yield
of the
leading order loop diagrams ($q^3$) is approximately 
$\frac{2}{3}$ of the total 2\Piz{} strength.
This fact makes this channel unique, because unlike in other
channels where the loops are adding some contribution to the dominant tree
graphs, here they dominate.
Consequently the 2\Piz{} channel provides a very sensitive method
to study these loop contributions to ChPT. In \cite{bernard:2pi0thres},
the following prediction for the near threshold cross section was given:
\begin{eqnarray}
    \sigma_{tot}(\Egam) = && \mbox{0.6 nb} \biggr(
    \frac{\Egam{} - \Egam^{thr}}{\mbox{10 MeV}}\biggr)^2
    \label{eq:sigma}
\end{eqnarray}
where \Egam{} denotes the photon beam energy and $\Egam^{thr}$ is 
the production
threshold of 308.8 MeV. 
Actually, the uncertainty of the 
coupling of the \Roper{} to the s-wave $\pi \pi$ channel was 
a limiting factor
for the accuracy of the ChPT calculation \cite{bernard:2pi0thres}.
For the most extreme case of this coupling, 
an upper limit for this cross section was deduced
by increasing the constant
in Eq.~\ref{eq:sigma} from 0.6 nb to 0.9 nb. 

To complete the overview of theoretical calculations of the reaction \Rpipi{}
close to threshold, it is noted that this channel
is also described in a recent version of the
Gomez Tejedor-Oset model \cite{tejedor:pipi}.
This model is based on a set of tree level diagrams
including pions, nucleons and nucleonic resonances. In a recent work,
particular emphasis was put on the rescattering of pions in
the isospin $I$=0 channel \cite{roca:pipi}.
Double pion photoproduction via the $\Delta$ Kroll-Rudermann term is not
possible for the 2\Piz{} final state.
In the case of a $\pi^- \pi^+$ Kroll-Rudermann
term, the charged pions can rescatter into two neutral pions generating
dynamically a $\pi \pi$ loop. This effect nearly doubles the cross
section in the threshold region and is regarded by the authors as being
reminiscent of the explicit chiral loop effect described above.
Nevertheless, the cross section calculated with this model is
significantly smaller than the ChPT prediction.

In the past, two measurements of
the reaction \Rpipi{} below 450~MeV beam energy have been carried out 
\cite{haerter:pipi,wolf:pipi}.
The second experiment showed an improvement in statistics by
almost a factor 30.
Nevertheless, in the threshold region the cross
section still suffered from large statistical uncertainties (see
Fig.~\ref{fig:cross}).

In contrast to previous analyses, we did not
extract the cross section from events in which 
only three of the four decay photons were detected.
Such analyses are not kinematically overdetermined 
and, close to threshold in particular, the extracted cross section 
can be slightly contaminated by the reaction $\gamma
p \rightarrow \Piz{} \gamma p$, which has recently been measured 
\cite{kottu:mdm_prl}.
This present measurement of the 2\Piz{} photoproduction at threshold
is the first
for which comparison to theoretical calculations is conclusive.

\section{Experimental Setup and Data Analysis}
\label{experiment}

The reaction \Rpipi{} was measured
at the Mainz Microtron (MAMI) electron accelerator 
\cite{walcher:mami,ahrens:mami} using the
Glasgow tagged photon facility \cite{anthony:tagger,hall:tagger}
and the photon spectrometer TAPS \cite{novotny:taps,gabler:response}.
A quasi-monochromatic photon beam was produced by
bremsstrahlung tagging, in which the photon energy is determined by the
difference between the energies of the incident electron and the residual
electron following bremsstrahlung because the energy transfer to the atomic
nuclei of the radiator foil is negligible.
The photon energy covered the range 285--820 MeV with an average energy
resolution of 2 MeV. The photon flux was of the order of 
$0.5$ $\mbox{MHz} \mbox{MeV}^{-1}$ at photon energies of 300~MeV.
The TAPS detector consisted of six blocks each with 62 hexagonally shaped
BaF$_2$ crystals arranged in an 8$\times$8 matrix and a forward wall
with 138 BaF$_2$ crystals arranged in a 11$\times$14 rectangle.
Each crystal is 250~mm long with an inner diameter of 59~mm.
The six blocks were
located in a horizontal plane around the target at angles of
$\pm$54$^{\circ}$, $\pm$103$^{\circ}$ and $\pm$153$^{\circ}$ with
respect to the beam axis. Their distance to the target was 55~cm and the
distance of the forward wall was 60~cm.
This setup covered $\approx$40\% of the full solid angle.
All BaF$_2$ modules were equipped with 5~mm thick scintillation plastic dE/dx 
detectors to allow the
identification of charged particles.
The liquid hydrogen target was 10~cm long with a diameter of 3~cm.
Further details of the experimental setup can be found in ref.
\cite{kottu:mdm_erice}.

\begin{figure}
   \includegraphics[width=0.5\columnwidth]{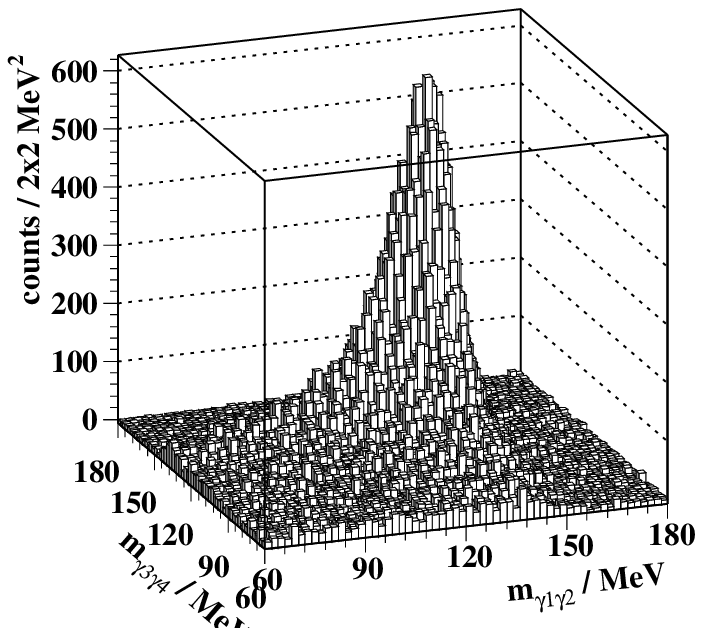}
   \includegraphics[width=0.5\columnwidth]{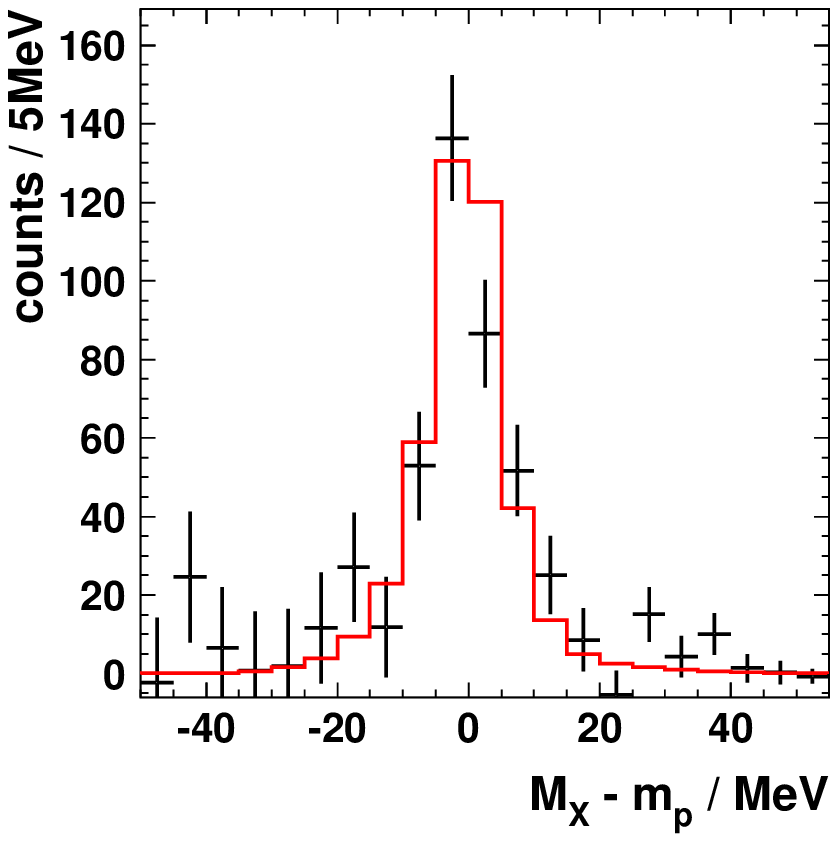}
\caption{Left: Two photon invariant masses $m_{\gamma 1 \gamma 2}$ vs
   $m_{\gamma 3 \gamma 4}$. Right: Missing mass $M_X - m_p$ derived from
   two detected \Piz{} mesons for
incident beam energies \Egam{} $\leq$ 400 MeV (symbols with errors: data,
histogram: GEANT simulation).}
\label{fig:kin}
\end{figure}

The \Rpipi{} reaction channel was identified by measuring
the 4-momenta of the two \Piz{} mesons, whereas the proton was not
detected. For a three-body final state this provides
kinematical overdetermination and hence an unambiguous identification of this
reaction channel.
The \Piz{} mesons were detected via their two-photon decay channel and
identified in a standard invariant mass analysis from the measured photon
momenta.
The four photons of an event can be arranged in three different 
combinations to form two
2-photon invariant masses (compare Fig.~\ref{fig:kin}).
For an acceptable ($\gamma , \Piz \Piz{}$) event, 
one of these combinations was required to fullfil the
condition, $110 MeV < m_{\gamma \gamma} < 150 MeV$,
for both of the 2-photon invariant masses.
In addition the mass of the missing proton was calculated from the beam
energy $E_{beam}$, target mass $m_p$ and the energies $E_{\Piz{}}$ 
and momenta $\vec{p}_{\Piz{}}$ of the pions via:
\begin{eqnarray}
    M_{X}^2 = && ((E_{\Piz_1}+E_{\Piz_2})-(E_{beam}+m_{p}))^2 
     - ((\vec{p}_{\Piz_1}+\vec{p}_{\Piz_2})-(\vec{p}_{beam}))^2
    \label{eq:mmiss}
\end{eqnarray}
The resulting distribution is
shown in Fig.~\ref{fig:kin}. In case of the reaction \Rpipi{} the missing
mass must be equal to the mass of the (undetected) proton $m_p$.
A Monte Carlo simulation 
of the 2\Piz{} reaction using GEANT3 \cite{geant} 
reproduces the lineshape of the measured data. A cut corresponding to a $\pm
2 \sigma$ width of the simulated lineshape has been applied to select the
events of interest.
Random time coincidences between the 
TAPS detector and the tagging spectrometer
were subtracted, using 
events outside the prompt time coincidence window\cite{hall:tagger}.
No systematical errors are introduced by this method and the statistical 
errors introduced by the subtraction are included in the
errors presented in this work.

The cross section was deduced from the rate of
the 2\Piz{} events, the number of hydrogen atoms
per cm$^2$, the photon beam flux, the branching ratio of the 
\Piz{} decay into two
photons, and the detector and analysis efficiency.
The intensity of the photon beam was determined by counting the
scattered electrons in the tagger focal plane and measuring the
loss of photon intensity due to collimation 
with a 100\%-efficient BGO detector 
which was moved into the
photon beam at reduced intensity. 
The geometrical detector acceptance and
analysis efficiency due to cuts and thresholds 
were obtained using the GEANT3 code and an event
generator producing distributions of the final state particles
according to phase space. The acceptance of the detector setup was studied by
examining independently 
a grid of the four degrees of freedom for this three body 
reaction (azimuthal
symmetry of the reaction was assumed). In a grid of total 1024 bins 
no acceptance holes were 
found for the beam energy range presented in this paper.
The average value for the detection 
efficiency is 1.0\%. The systematic errors are estimated to be 6\% and
include uncertainties of the beam flux, the
target length and the efficiency determination.
All results presented in this work are acceptance corrected and absolutely
normalized.

\section{Results and Discussion}
\label{}

\begin{figure*}
  \includegraphics[width=0.99\columnwidth]{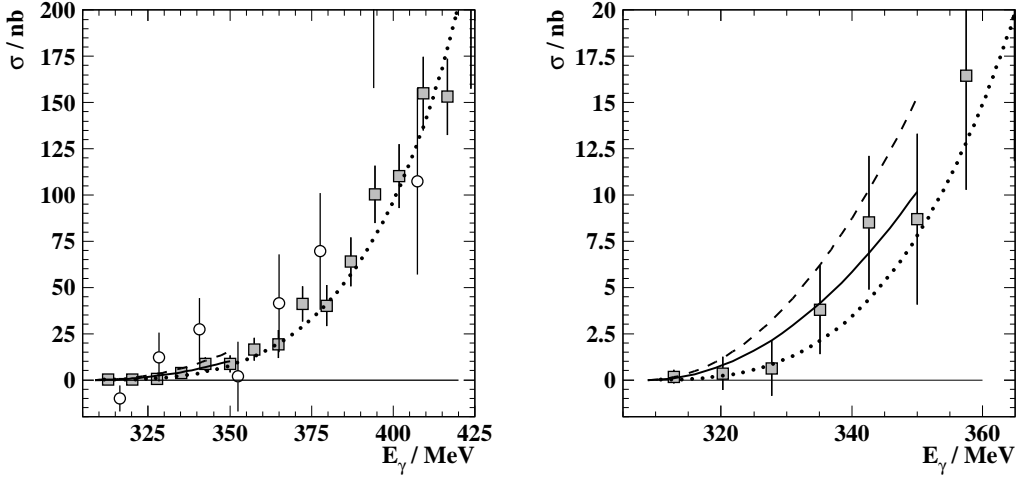}
  \caption{Total cross section for the reaction \Rpipi{} (grey squares) at
  threshold in comparison with a previous
  experiment \cite{wolf:pipi} (open circles) for incident energies up to
  360~MeV (right) and 425~MeV (left), respectively. The error bars denote the
  statistical error.
  The prediction
  of the ChPT calculation \cite{bernard:2pi0thres} is shown (solid curve)
  together with its upper limit (dashed curve) and the prediction of Ref.
  \cite{roca:pipi} (dotted curve).
 }
  \label{fig:cross}
\end{figure*}

The measured total cross section for the
reaction \Rpipi{} is shown in Fig.~\ref{fig:cross} as a function of the
incident photon beam energy. The results are compared to
a previous experiment \cite{wolf:pipi}. 
The two experiments
are consistent within the rather large errors of the previous work.

The prediction of ChPT \cite{bernard:2pi0thres} is plotted
up to 40 MeV above the production threshold. The overall shape as well as the 
absolute magnitude are in agreement with the data.
Furthermore, the ChPT calculation using the upper limit 
for the \Roper{} coupling to the s-wave $\pi \pi$ channel,
can be excluded.
In the future, the present data might
be exploited to provide a better constraint on this coupling.
Additionally, the cross section is compared to the calculation with
the chiral unitary model \cite{roca:pipi}, which especially at threshold
predicts a smaller cross section.
The data show good agreement with both calculations.

\begin{figure*}
  \includegraphics{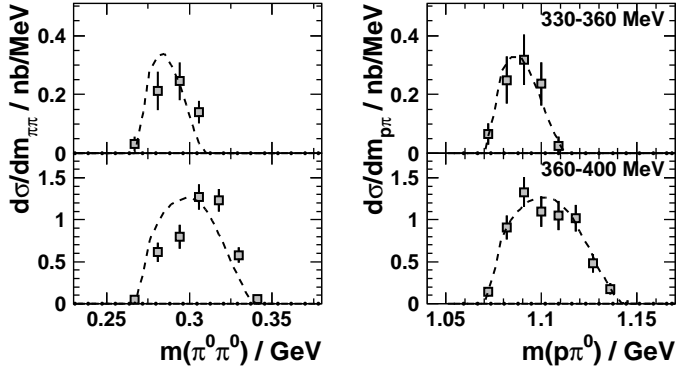}
  \caption{
  Invariant mass of \Piz \Piz{} and \Piz $p$  for different bins of 
  beam energy (full squares).
  The dashed curve shows 3 body phase space.
  The beam energy range in the
 upper panel is 330-360~MeV and in the lower panel 360-400~MeV.
 }
  \label{fig:dalitz1}
\end{figure*}

Fig.~\ref{fig:dalitz1} shows the
invariant masses of the $\Piz \Piz$ and $p \Piz{}$ systems in the incident
energy beam ranges of 330-360~MeV and 360-400~MeV.
The $p \Piz$ mass is
consistent with a three body phase space distribution, whereas the $\Piz
\Piz$ mass deviates slightly already for the energy bin of 330-360~MeV
from the phase space distributions and shows a trend towards higher
invariant masses. The Valencia chiral unitary model \cite{roca:pipi} 
explains that $m_{\Piz \Piz}$ 
distributions skewed to higher invariant masses can
arise from the interference of the isospin I=0 and I=2
$\Piz \Piz$ amplitudes.

\begin{figure*}
  \includegraphics{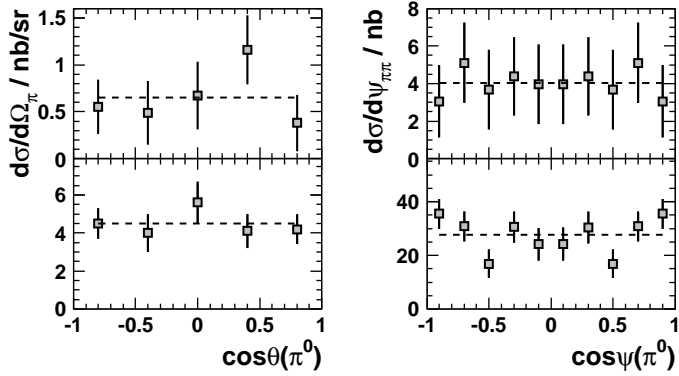}
  \caption{Angular distributions of the center of mass polar
  angle $\theta_{\Piz{}}$ (left panel) 
  and the angle $\psi_{\Piz}$ (right panel) 
  between the proton and the 
  two-pions in the $\Piz \Piz$ 
  rest frame (Gottfried Jackson system).
  The dashed curve shows 3 body phase space. The beam energy range in the
 upper panel is 330-360~MeV and in the lower panel 360-400~MeV.
 }
  \label{fig:dalitz2}
\end{figure*}

Two angular distributions are 
depicted in Fig.~\ref{fig:dalitz2}. The polar angle $\theta_{\Piz}$ 
of the \Piz{} mesons in the overall center of mass frame is consistent with
an isotropic distribution. The same holds for the
angle between the $\Piz$ mesons 
$\psi_{\Piz}$ and the proton
in the frame where the $\Piz \Piz$ system is at rest (Gottfried Jackson
system). Due to the indistinguishability of the two \Piz{} mesons, the
distribution of $\psi_{\Piz}$ shows a symmetry around $90^\circ$.
The isotropy with respect to the $\psi_{\Piz}$ angle in both energy ranges
indicates, that the \Piz{} mesons are dominantly in an
s-wave state.

In summary, we have measured the total cross section at threshold 
for the reaction \Rpipi{}. This experiment was motivated as a test of
a ChPT calculation \cite{bernard:2pi0thres} which shows an unexpectedly high
2\Piz{} rate compared to final states including charged pions. This fact is
attributed to a dominant contribution of pion loops which appear in leading
(non vanishing) order. This prediction is in agreement with our measured data.
Furthermore, the upper limit quoted
for this prediction can be excluded. In future it might be possible to 
exploit the present data to give a better
constraint on the \Roper{} to s-wave $\pi \pi$ coupling. Secondly,
the data are also compared to a prediction \cite{roca:pipi},
where pion loops are dynamically generated.
Especially close to threshold, the two predicted cross sections 
differ significantly.
Although the present data are of much superior statistical quality than
previous measurements, the precision is still not good enough to discriminate 
between these two models. The observed angular distribution 
show that the \Piz{} mesons are dominantly emitted in an s-wave state. 

We thank the accelerator group of MAMI 
as well as many other scientists and technicians of the Institut fuer
Kernphysik at the University of Mainz for their outstanding support.
This work is supported by DFG Schwerpunktprogramm:
"Untersuchung der hadronischen Struktur von Nukleonen und Kernen mit
elektromagnetischen Sonden", SFB221, SFB443, 
the UK Engineering and Physical Sciences Research 
Council and Schweizerischer Nationalfond.



\bibliographystyle{paper}
\bibliography{kotulla_2pi0thres.bbl}





\end{document}